\newif\iffigs\figstrue

\documentstyle[12pt]{article}
\iffigs
  \input epsf
\else
\fi

\batchmode
  \newfont{\footscrfont}{rsfs10}
  \newfont{\footbbbfont}{msbm10}
\errorstopmode

\newif\ifscrf\scrftrue
\ifx\footscrfont\nullfont
  \scrffalse
\fi

\newif\ifamsf\amsftrue
\ifx\footbbbfont\nullfont
  \amsffalse
\fi

\def\ppnumber{\vbox{\baselineskip16pt\hbox{CLNS-95/1369}
\hbox{LMU--TPW 95--16}
\hbox{hep-th/9510234}}}
\def\ppdate{October 1995}
\def\pplogo{\vbox{\kern-\headheight\kern -33pt
\halign{##&##\hfil\cr&{
\ppnumber}\cr\rule{0pt}{2.5ex}&\ppdate\cr}
}}

\makeatletter
\date{}
\def\dedicatory#1{\def\@date{\normalsize\it#1}}
\def\subjclass#1{\def\@thefnmark{}\@footnotetext{1991
    {\it Mathematics Subject Classification.} #1}}
\def\keywords#1{\def\@thefnmark{}\@footnotetext{
    {\it Key words and phrases.} #1}}

\def\ps@firstpage{\ps@empty \def\@oddhead{\hss\pplogo}%
  \let\@evenhead\@oddhead 
}
\def\maketitle{\par
 \begingroup
 \def\thefootnote{\fnsymbol{footnote}}
 \def\@makefnmark{\hbox
 to 0pt{$^{\@thefnmark}$\hss}}
 \if@twocolumn
 \twocolumn[\@maketitle]
 \else \newpage
 \global\@topnum\z@ \@maketitle \fi\thispagestyle{firstpage}\@thanks
 \endgroup
 \setcounter{footnote}{0}
 \let\maketitle\relax
 \let\@maketitle\relax
 \gdef\@thanks{}\gdef\@author{}\gdef\@title{}\let\thanks\relax}

\def\abstract{\if@twocolumn
\section*{Abstract}
\else \small
\begin{center}
{\bf ABSTRACT}
\end{center}
\quotation
\fi}

\newtheorem{theorem}{Theorem}

\def\thebibliography#1{\section*{References\@mkboth
 {REFERENCES}{REFERENCES}}\small\list
 {[\arabic{enumi}]}{\settowidth\labelwidth{[#1]}\leftmargin\labelwidth
 \advance\leftmargin\labelsep
 \usecounter{enumi}}
 \def\newblock{\hskip .11em plus .33em minus .07em}
 \sloppy\clubpenalty4000\widowpenalty4000
 \sfcode`\.=1000\relax}

\newif\iffn\fnfalse

\@ifundefined{reset@font}{\let\reset@font\empty}{} 
\long\def\@footnotetext#1{\insert\footins{\reset@font\footnotesize
    \interlinepenalty\interfootnotelinepenalty
    \splittopskip\footnotesep
    \splitmaxdepth \dp\strutbox \floatingpenalty \@MM
    \hsize\columnwidth \@parboxrestore
   \edef\@currentlabel{\csname p@footnote\endcsname\@thefnmark}\@makefntext
    {\rule{\z@}{\footnotesep}\ignorespaces
      \fntrue#1\fnfalse\strut}}}

\newcount\@tempcntc
\def\@citex[#1]#2{\if@filesw\immediate\write\@auxout{\string\citation{#2}}\fi
  \@tempcnta\z@\@tempcntb\m@ne\def\@citea{}\@cite{\@for\@citeb:=#2\do
    {\@ifundefined
       {b@\@citeb}{\@citeo\@tempcntb\m@ne\@citea
        \def\@citea{,\penalty\@m\ }{\bf ?}\@warning
       {Citation `\@citeb' on page \thepage \space undefined}}%
    {\setbox\z@\hbox{\global\@tempcntc0\csname b@\@citeb\endcsname\relax}%
     \ifnum\@tempcntc=\z@ \@citeo\@tempcntb\m@ne
       \@citea\def\@citea{,\penalty\@m}
       \hbox{\csname b@\@citeb\endcsname}%
     \else
      \advance\@tempcntb\@ne
      \ifnum\@tempcntb=\@tempcntc
      \else\advance\@tempcntb\m@ne\@citeo
      \@tempcnta\@tempcntc\@tempcntb\@tempcntc\fi\fi}}\@citeo}{#1}}

\def\@citeo{\ifnum\@tempcnta>\@tempcntb\else\@citea
  \def\@citea{,\penalty\@m}%
  \ifnum\@tempcnta=\@tempcntb\the\@tempcnta\else
   {\advance\@tempcnta\@ne\ifnum\@tempcnta=\@tempcntb \else
\def\@citea{--}\fi
    \advance\@tempcnta\m@ne\the\@tempcnta\@citea\the\@tempcntb}\fi\fi}

\makeatother

\ifamsf
  \newfont{\bigbbbfont}{msbm10 scaled\magstep2}
  \newfont{\bbbfont}{msbm10 scaled\magstep1}  
  \newfont{\smallbbbfont}{msbm8}
  \newfont{\tinybbbfont}{msbm6}
  \newfont{\smallfootbbbfont}{msbm7}
  \newfont{\tinyfootbbbfont}{msbm5}
\fi

\ifscrf
  \newfont{\scrfont}{rsfs10 scaled\magstep1}  
  \newfont{\smallscrfont}{rsfs7}
  \newfont{\tinyscrfont}{rsfs7}
  \newfont{\smallfootscrfont}{rsfs7}
  \newfont{\tinyfootscrfont}{rsfs7}
\fi

\ifamsf
  \newcommand{\Bbb}[1]{\iffn
      \mathchoice{\mbox{\footbbbfont #1}}{\mbox{\footbbbfont #1}}
      {\mbox{\smallfootbbbfont #1}}{\mbox{\tinyfootbbbfont #1}}\else
      \mathchoice{\mbox{\bbbfont #1}}{\mbox{\bbbfont #1}}
      {\mbox{\smallbbbfont #1}}{\mbox{\tinybbbfont #1}}\fi}
\else
  \def\bigbbbfont{\bf}
  \def\Bbb{\bf}
\fi

\ifscrf
  \newcommand{\Scr}[1]{\iffn
    \mathchoice{\mbox{\footscrfont #1}}{\mbox{\footscrfont #1}}
    {\mbox{\smallfootscrfont #1}}{\mbox{\tinyfootscrfont #1}}\else
    \mathchoice{\mbox{\scrfont #1}}{\mbox{\scrfont #1}}
    {\mbox{\smallscrfont #1}}{\mbox{\tinyscrfont #1}}\fi}
\else
  \def\Scr{\cal}
\fi

\def\operatorname#1{\mathop{\rm #1}\nolimits}
\def\C{{\Bbb C}}

\def\P{{\Bbb P}}
\def\Q{{\Bbb Q}}
\def\R{{\Bbb R}}
\def\Z{{\Bbb Z}}

\def\Area{\operatorname{Area}}

\def\rank{\operatorname{rank}}

\def\opeq#1{\advance\lineskip#1 \advance\baselineskip#1
        \advance\lineskiplimit#1}

\def\eqalign#1{\null\,\vcenter{\opeq{2.5\jot}\mathsurround=0pt
        \everycr={}\tabskip=0pt
        \halign{\strut\hfil$\displaystyle{##}$&$\displaystyle{{}##}$\hfil
        \crcr#1\crcr}}\,\null}

\def\sm{$\sigma$-model}
\def\nlsm{non-linear \sm}

\def\CY{Calabi--Yau}

\def\cM{{\Scr M}}

\def\cD{{\Scr D}}

\def\cMc{{\hfuzz=100cm\hbox to 0pt{$\;\overline{\phantom{X}}$}\cM}}
\def\barcD{{\hfuzz=100cm\hbox to 0pt{$\;\overline{\phantom{X}}$}\cD}}

\def\ff#1#2{{\textstyle\frac{#1}{#2}}}

\ifamsf

\else

\fi

\begin{document}
\setcounter{page}0
\title{\LARGE On the Ubiquity of K3 Fibrations\\
  in String Duality\\[10mm]}
\author{
Paul S. Aspinwall\\[0.7cm]
\normalsize F.R.~Newman Lab.~of Nuclear Studies,\\
\normalsize Cornell University,\\
\normalsize Ithaca, NY 14853\\[10mm]
Jan Louis\\[0.7cm]
\normalsize Sektion Physik, \\
\normalsize Universit\"at M\"unchen, \\
\normalsize Theresienstrasse 37, \\
\normalsize D-80333 M\"unchen, \\
\normalsize Germany\\[5mm]
}

{\hfuzz=10cm\maketitle}

\def\Large{\large}
\def\LARGE{\large\bf}

\vskip 1cm

\begin{abstract}

We consider the general case of $N=2$ dual pairs of type IIA/heterotic
string theories in four dimensions. We show that if the type IIA
string in this pair can be viewed as having been compactified on a
Calabi-Yau manifold in the usual way then this manifold must be of the
form of a K3 fibration. We also see how the bound on the rank of the
gauge group of the perturbative heterotic string has a natural
interpretation on the type IIA side.

\end{abstract}

\vfil\break


\section{Introduction}

Recently, the quantum properties of $N=2$ supersymmetric field
theories have been under active investigation.  In particular, Seiberg
and Witten \cite{SW:I} analyzed an asymptotically free $SU(2)$
Yang-Mills theory and determined the leading contribution of the low
energy effective action for all values of the gauge coupling.  It
turns out that in the strong coupling region no massless $SU(2)$ gauge
bosons exist; instead magnetic monopoles become massless and the
effective low energy theory is best described by a weakly coupled dual
$U(1)$ magnetic gauge theory.

These developments lead to intensive attempts to uncover the same
physical phenomenon in string theory and study its gravitational and
``stringy'' generalizations.  It has been conjectured that for strongly
coupled heterotic $N=2$ vacua the ``dual'' description is provided by
weakly coupled $N=2$ vacua of type II strings.  This conjecture is
supported by a number of concrete ``pairs'' of dual vacua
\cite{KV:N=2,FHSV:N=2,VW:pairs,SV:pairs} where the perturbative
effective action of the
heterotic vacuum (at least, for the vector multiplets) has been
matched with the dual type II vacuum in an appropriate expansion
\cite{KV:N=2,KLT:limit,KLM:K3f,AGN:N=2,AP:limit}.
Furthermore Seiberg--Witten theory should appear
in this context \cite{Bill:limit,CLM:limit} and can be recovered in
the $\alpha' \to 0$ limit \cite{KKL:limit}.

These encouraging results make it interesting to go beyond a model by
model analysis and study more generally the properties of the
conjectured duality. However, for an arbitrary heterotic string
vacuum it is presently unknown how to construct the dual partner.
One might suspect that this duality in 4 dimensions
is a consequence of string-string duality
\cite{HT:unity,Duff:S,W:dyn,Sen:sol}
in six dimensions and the dual pairs of
\cite{FHSV:N=2,me:flower} are indeed constructed in this way. It is
not clear how to show the connection for the pairs of \cite{KV:N=2}
however. Initial steps in this latter case were taken in
\cite{VW:pairs} where it was noticed
that both the type II string and
the heterotic string could be ``fibred'' over a $\P^1$. That is, the
\CY\ manifold, $X$, on which the type IIA string was compactified
could be written as a fibre bundle with base $\P^1$ and generic fibre
a K3 surface where the heterotic string was compactified on K3$\times
T^2$ which can be written as a bundle with base $\P^1$ and generic
fibre $T^4$. It is then tempting to try to use string-string duality
fibre-wise to map the type IIA string on the K3 surface into a
heterotic string on $T^4$.

The relevance  of K3-fibrations was first noticed in
\cite{KLM:K3f} where it was shown that such a structure
fits nicely with properties of the dual heterotic string compactified on
K3$\times T^2$.
(Various related aspects of K3-fibrations were also discussed
 in \cite{LY:K3f,AFIQ:chains}.)
In this paper we will take this point  further and
argue that the appearance of the K3-fibration is actually unavoidable.
The argument will borrow results from the ``phase'' picture of the
moduli space of $N=2$ conformal field theories
\cite{W:phase,AGM:II}. In particular, we need to understand clearly
what we mean by an object that is recognizably a heterotic string and
an object that is recognizably a type IIA string theory compactified
on a \CY\ manifold. In the latter case, we know from the phase picture
that we need to identify a ``\CY'' phase of the moduli space and we
will see that we must also identify a ``weakly-coupled'' phase in the
heterotic string moduli space.

The phase picture allowed a complete resolution of the question of
\CY\ manifolds without mirror partners \cite{AG:gmi}. We will see that
exactly the same considerations apply to heterotic strings without dual
type II partners and {\em visa versa}. If we demand that we do have a
dual pair in which the weakly-coupled phase of a heterotic string maps
into the \CY\ phase of a type IIA string then we will see that this
\CY\ manifold must have a K3-fibration.

One bound which appears naturally on the heterotic side
is on the rank of the gauge group. One can use conformal field theory
considerations to show that it cannot exceed 24. It has been troubling
that no such bound is known on the type II side. We will see how
K3-fibrations
give a natural interpretation to this effect
although it appears that this bound can be broken by nonperturbative
effects on the heterotic side.


\section{Perturbative $N=2$ Heterotic String Vacua}     \label{s:het}

\def\Mh{\cM_{\rm h}}
\def\Mv{\cM_{\rm v}}
\def\nh{n_{\rm h}}
\def\nv{n_{\rm v}}
\def\cb{\bar{c}}
\def\Re{{\rm Re}}
\def\Ghet{{G_{\rm het}}}
\def\GII{{ G_{ II}}}
\def\GNA{G^\prime}
\def\del{\partial}
\def\delbar{\bar\partial}
\def\a{\alpha} \def\b{\beta} \def\d{\delta}
\def\e{\epsilon} \def\c{\gamma}

Let us start by summarizing the generic properties of perturbative
heterotic $N=2$ vacua.  The heterotic string is based on a conformal
field theory with central charge $(c,\bar c)=(26,15)$.  Out of this
total central charge a $(c,\bar c)=(4,6)$ ``block'' is used to build the
four-dimensional space-time while the left over $(c,\bar c)=(22,9)$
conformal field theory is only further constrained by the amount of
space-time supersymmetry and the size of the gauge group.  $N=2$
space-time supersymmetry requires that the right moving internal $\bar
c =9$ conformal field theory splits into a free complex boson (of
central charge $\bar c=3$) and a $\cb =6$ conformal field theory with
$N=4$ world-sheet supersymmetry \cite{BD:susy}.  The gauge group $\Ghet$
arises from the
left moving $c=22$ conformal field theory and the free $\cb =3$ boson.
The latter gives rise
to two Abelian $U(1)$ gauge bosons which are identified with
the graviphoton and the vector partner of the dilaton.
The $c=22$ conformal field theory generates an arbitrary (possibly
non-Abelian) gauge
group $\GNA$ of maximal rank 22.
Therefore $\Ghet = \GNA \times [U(1)]^2$ and
the rank of $\Ghet$ is bounded by
\begin{equation}
2 \le {\rm rank}(\Ghet) \le 24.
     \label{eq:hetrank}
\end{equation}

The supersymmetric partners of the gauge bosons are two gauginos and a
complex scalar all in the adjoint representation of $\Ghet$; they form
what is called an $N=2$ vector multiplet.  The matter fields usually
transform in the fundamental representation of $\Ghet$ and they reside
in $N=2$ hypermultiplets which contain two Weyl fermions and four real
scalars.  All string vacua also feature gauge neutral scalars termed
moduli which are exactly flat directions of the effective potential and
which parameterize the perturbative degeneracy of a given vacuum
family. Such moduli arise as the scalars of either vector or
hypermultiplets and they can be thought of as parameters taking values
in a moduli space.  $N=2$ supersymmetry implies that the moduli space $\cM$
is always locally a product \cite{deW:prod}
\begin{equation}
 \cM = \Mv\times \Mh\ ,
     \label{eq:modulispace}
\end{equation}
where $\Mv$ ($\Mh$) is spanned by
the scalars in the vector multiplets (hypermultiplets).

The  scalar fields $T^i, i=1, \ldots, r \equiv {\rm rank}(\GNA)  \le 22$
in the Cartan subalgebra of $\GNA$
are flat directions of the effective potential.
If all $T^i$'s are  turned on
(a generic point in the moduli space)
$\GNA$ is broken to its maximal Abelian subgroup
(which is at most $[U(1)]^{22}$)
while at special points in the moduli space this gauge symmetry can be
enhanced to a non-Abelian group.

In addition to  the $T^i$ moduli there is also always
the string dilaton present in the massless spectrum.
Its vacuum expectation value  determines
the string coupling constant and   organizes the string
perturbation theory.
Together with the dual axion of an antisymmetric
tensor field it forms  a complex scalar $S$ of
an Abelian  $U(1)$ vector multiplet\footnote{
 The $U(1)$ gauge bosons arises in the $\cb=3$ block
 of the conformal field theory.}
and in the standard heterotic normalization one defines
\begin{equation}
S = {1\over g^{2}} - i {\theta\over 8\pi^2}\ ,
     \label{eq:Sdef}
\end{equation}
such that  large $S$ corresponds to  weak coupling.
With this convention
the Peccei--Quinn symmetry associated with the
axion is given by   a shift
\begin{equation}
\theta \to \theta + 2\pi r\ , \qquad
S \to S - {i r \over 4\pi}\ ,
     \label{eq:PQshift}
\end{equation}
where $r$ is a continuous parameter in string perturbation
theory. However, once nonperturbative corrections are
taken into account this continuous symmetry is broken down
to a discrete subgroup and then $r$ must be an integer.

The couplings of the vector multiplets (the K\"ahler potential and the
gauge couplings) are entirely encoded in a holomorphic prepotential
$F_0$.  Due to the non-renormalization theorem of $N=2$ supersymmetry
$F_0$ does not receive quantum
corrections beyond one-loop in
perturbation theory but can (and is) corrected nonperturbatively.
Using the fact that the dilaton counts string loops, $F_0$ can be
expanded for large $S$ (weak coupling) according to
\begin{equation}
F_0 = F_0^{(0)}(S,T^i) + F_0^{(1)} (T^i) + F_0^{(np)}(S,T^i)\ ,
     \label{eq:Fhet}
\end{equation}
where $F_0^{(0)}$ $\left(F_0^{(1)}\right)$
 is the tree level
(one-loop) contribution while
$F_0^{(np)}$  denotes nonperturbative corrections.
 The universal couplings of the dilaton and the continuous
Peccei--Quinn symmetry of the axion completely determine $F_0^{(0)}$
for all heterotic string vacua to be
\cite{FGKP:eff,FvP:Ka,CDFvP:,DKLL:}
(using the conventions of \cite{DKLL:})
\begin{equation}
F_0^{(0)}  = - \gamma_{ij} T^i T^j S \ ,
\qquad
\gamma_{ij} = {\rm diag} (+,-,\ldots,-)\ ,
     \label{eq:Fhettree}
\end{equation}
where $S$ and $T^i$ are the ``special'' $N=2$ coordinates.
$F_0^{(1)}(T^i)$ is necessarily
$S$-independent.
The tree level prepotential given in (\ref{eq:Fhettree}) corresponds
to the K\"ahler potential
\begin{equation}
K = - \log(S +\bar S)
-\log\gamma_{ij}(T^i + \bar{T}^i) (T^j + \bar{T}^j)\ ,
     \label{eq:Kahler}
\end{equation}
which is the K\"ahler potential of the homogeneous space
\begin{equation}
\Mv^{\,0} = {SU(1,1)\over U(1)} \times {SO(2,r) \over SO(2) \times SO(r)},
     \label{eq:Mhet}
\end{equation}
up to discrete identifications. Note that
(\ref{eq:Mhet}) is subject to string loop corrections and, in general,
is not isomorphic to $\Mv$.

So far we concentrated on the lowest order (two derivative)
couplings in the effective action.
There is a special class of higher derivative curvature couplings
which arise from chiral integrals in $N=2$ superspace and therefore
are also governed by holomorphic functions $F_n(S,T^i)$ of the vector
multiplets.  These are couplings of the form $g^{-2}_n R^2 G^{2n-2}$
where $R$ is the (anti-self-dual)  Riemann tensor, $G$ is the field strength of
the
graviphoton and the couplings $g_n$ obey \cite{BCOV:ell,BCOV:big,AGNT:top}
\footnote{We slightly change notation here. In this paper $F_n$ always
is a holomorphic quantity while  $g_n$ is the  non-harmonic coupling.}
\begin{equation}
g^{-2}_n  = \Re F_n(S,T^i) + {\cal A}_n \ .
     \label{eq:gdef}
\end{equation}
At the string tree level ${\cal A}_n =0$ holds and therefore
$g^{-2}_n$ is a harmonic function of the vector moduli.  However, at
one-loop  $g^{-2}_n$ develops
a holomorphic anomaly ${\cal A}_n$.  For $n=1$ one finds
\cite{CO:Ka,IL:anom,AGN:modcII,KLT:limit}
\begin{equation}
\del_{i} \bar{\del}_{\bar\jmath}\ g^{-2}_1
= {b\over 16 \pi^2}\ \del_i \bar{\del}_{\bar\jmath} K \ ,\qquad
b = 2(\nh - \nv + 23)\ ,
     \label{eq:anomaly}
\end{equation}
where $K$ is the K\"ahler potential defined in (\ref{eq:Kahler});
$\nh$ ($\nv$)
counts the   number of hypermultiplets (vector multiplets).
 Similar to (\ref{eq:Fhet}) the
holomorphic $F_n$'s can be expanded
in the dilaton according to\footnote{We thank B.~de Wit for discussions
of this point.}
\begin{equation}
F_n = F_n^{(0)}(S,T^i) + F_n^{(1)} (T^i) + F_n^{(np)}(S,T^i)\ .
     \label{eq:Fnhet}
\end{equation}
Again $ F_n^{(1)} (T^i)$ is dilaton independent  and
the perturbative Peccei--Quinn symmetry of the axion
determines
\begin{equation}
F_1^{(0)} =  24\, S\ , \qquad
F_{n>1}^{(0)}  = 0\ ,
     \label{eq:Fnhettree}
\end{equation}
where the normalization of $F_1^{(0)}$ is fixed by the
normalization of (\ref{eq:anomaly}).


\section{The Dual Type IIA String}   \label{s:wII}

Type II string vacua are built from conformal field theories of central
charge $(c,\cb) = (15,15)$. The four-dimensional space-time degrees of
freedom arise from a $(c,\cb) = (6,6)$ block leaving $(c,\cb) = (9,9)$
as the internal conformal field theory.  The standard way to obtain
$N=2$ space-time supersymmetry is to demand a left-right symmetric
$N=2$ world-sheet supersymmetry of the internal $(c,\cb) = (9,9)$
conformal field theory.\footnote{We do not consider asymmetric
constructions \cite{DKV:4d} in this paper; some examples have recently been
discussed in \cite{VW:pairs,SV:pairs}.}
An example of such a conformal field theory is provided by the \nlsm\
on a Calabi-Yau manifold.
Let us consider a type IIA superstring compactified on a \CY\ manifold
$X$. Equivalently, we can compactify the type IIB string on $Y$, the
mirror of $X$.

The number of vector- and
hypermoduli is directly related to the Hodge numbers of the
Calabi--Yau compactification.  Specifically one finds $\nv
=h^{1,1}(X),\, \nh = h^{2,1}(X)+1$ for type IIA and $\nv =h^{2,1}(Y),\,
\nh = h^{1,1}(Y)+1$ for type IIB; in both cases the additional
hypermultiplet corresponds to the type II dilaton.
The gauge group $\GII$ always is a product of $(\nv + 1)$ Abelian $U(1)$
factors and therefore one has   in general
\begin{equation}
{\rm rank}(\GII) = \nv +1\ ,
     \label{eq:ttrank}
\end{equation}
where the ``$+1$''  counts the graviphoton.
The dilaton always is the member of a
hypermultiplet and thus (\ref{eq:modulispace}) implies that
for type II vacua  $\Mv$ is
independent on the dilaton and therefore
determined at the string tree level exactly with
no further corrections perturbatively or nonperturbatively.

Near the large radius limit of $X$, the position in the moduli space $\Mv$
is given by the complexified K\"ahler form on $X$ --- that is, an object of
the form $B+iJ\in H^2(X,\C)$, where we expand
\begin{equation}
  B+iJ = \sum_{\alpha=1}^{h^{1,1}} (B+iJ)_\alpha e_\alpha,
\end{equation}
so that $B_\alpha$ and $J_\alpha$ are real numbers and $e_\a$
represents a basis of $H^2(X,\Z)$ (we assume that $b_1(X)=0$).
As $H^2(X)$ is dual to $H_2(X)$ and $H_2(X)$ is dual to $H_4(X)$, we
have an isomorphism between $H^2(X)$ and $H_4(X)$. Using this
isomorphism, we can associate a divisor, $D_\a$, of $X$ to each $e_\a$.

In the large radius
limit (or large complex structure limit for type IIB) the holomorphic
prepotential has the generic structure\footnote{We
will use a dot to represent the intersection product between homology
classes and also the natural pairing between homology and
cohomology. We will also not distinguish between a divisor and its
homology class.}
\begin{equation}
F_0 = -\ff{i}{6}\sum_{\a,\b,\c}(D_\a\cdot D_\b\cdot D_\c)  t_\a t_\b
t_\c\ + \cdots+
        {\rm worldsheet\ instantons}\ ,
     \label{eq:Ftt}
\end{equation}
where $t_\a=(B+iJ)_\a$ denotes the moduli
of the vector multiplets in $N=2$ special coordinates and we have
omitted the \sm\ loop terms.

Recall \cite{DSWW:,AM:rat} that the instanton corrections in
(\ref{eq:Ftt}) come from rational curves within $X$. As these curves
get larger, the instanton effects become smaller. In order for
(\ref{eq:Ftt}) to make sense, we must assume that the sum produced by
the instantons is  convergent. If it is not, one can often replace
the \CY\ model by another theory (such as a Landau-Ginzburg theory) in
which some other instanton sum converges. This is regarded as the
underlying $N=2$ superconformal field theory
 being in another  ``phase'' \cite{W:phase,AGM:II}.
If we insist that we are in a \CY\ phase and thus that (\ref{eq:Ftt})
is a convergent series then we are demanding that all the rational
curves on $X$ are sufficiently large.

Now we are interested in identifying those Calabi--Yau compactification
which can serve as possible dual descriptions of heterotic vacua.
The idea is, that once all nonperturbative effects
have been taken into account, the heterotic theory and the type II
theory describe exactly the same physics in the infrared.
One test of the proposed duality is to check
that both theories have identical  moduli spaces.
However, in either case  we only know the respective weak coupling limits.
Nevertheless due to the product structure of the moduli space
and the fact that the dilaton resides in different super-multiplets in the two
theories
it is possible to compare the heterotic $\Mv$ in a weak
coupling expansion with the exact $\Mv$ of the type II vacuum.
In particular,  one of the vector moduli $t_\a$ in the type IIA
vacuum  must
be the image of the axion-dilaton pair of the heterotic string. Let us
denote this particular member by $t_s = (B+iJ)_s$
and let $D_s$ be the associated divisor.\footnote{Strictly speaking we
need to establish that $D_s$ is a ``$\Q$-Cartier'' divisor. This follows since
an irrational class would not be compatible with the Peccei--Quinn symmetry.}
As we have just said we can only compare the moduli spaces for large
$S$ and hence  $t_s$ also has to be large.
In this limit  the type IIA $B_s$-field
is a periodic variable and obeys  $B_s\to B_s+1$.
This is consistent with the
Peccei--Quinn symmetry $\theta\to\theta+2\pi$ of the
heterotic axion if we identify
\begin{equation}
  (B+iJ)_s = 4\pi i S.  \label{eq:ax-II}
\end{equation}
Using the fact that both the complexified K\"ahler form and the
axion-dilaton  are  special $N=2$ coordinates
(\ref{eq:ax-II}) is exact up to symplectic $Sp(2\nv+2,\Z)$
reparametrizations and not only valid at the large radius
limit.

The other vector multiplets will be denoted by $e_i$
(or equivalently $D_i$), for $i=1,\ldots,
\nv-1$.
Given our assumption that the weakly coupled phase
of the heterotic string maps into the \CY\ phase of the type IIA string
we can obtain from (\ref{eq:Fhettree}) and (\ref{eq:Ftt}) that

\begin{equation}
   D_s\cdot D_s\cdot D_s = 0 , \qquad\
   D_s\cdot D_s\cdot D_i = 0\quad\forall i.  \label{eq:ssi}
\end{equation}
$D_s\cdot D_s$ must be an element of the second homology of $X$. Since
(\ref{eq:ssi}) tells us that this element has zero intersection with
any 4-cycle, it must be trivial. That is,
\begin{equation}
  D_s\cdot D_s = 0.  \label{eq:num=1}
\end{equation}
We know that $D_s$ itself is not trivial as the associated modulus is
not trivial. In the language of algebraic geometry we have shown that
the ``numerical $D$-dimension'' of the divisor $D_s$ is one.

Next, let us analyze the positivity properties of $D_s$. We have
asserted that we are in the \CY\ phase of the type IIA string. From
the work of \cite{AGM:sd} it follows that we know something about
$J$. Very roughly, it means that $J$ lies within the K\"ahler cone of
$X$. That is, the volume of any algebraic subspace within $X$ is
measured by the K\"ahler form to be positive.\footnote{Actually, there are
quantum corrections to this statement and the walls of the
classical K\"ahler cone have to be moved in a little to make sure the
instanton sum remains finite. These corrections have no effect on our
argument here.}
Suppose we take a theory which is in the \CY\ phase and modify the
K\"ahler form as
\begin{equation}
  J \to J+\lambda e_s,
\end{equation}
where $\lambda$ is a large real number. In the heterotic language, this
amounts to increasing the value of the dilaton, i.e., making the
string more weakly coupled. Thus, the instanton sums will become more
convergent. If this is to have the same effect in the type IIA string,
then this change in the K\"ahler form must keep us inside the K\"ahler
cone of $X$. This amounts to the assertion that
\begin{equation}
  e_s\cdot C \geq 0          \label{eq:nef}
\end{equation}
for any algebraic curve, $C$, within $X$. In the language of algebraic
geometry, this tells us that $D_s$ is ``nef''.

Lastly we need one more property of $D_s$ which follows from
identifying the higher derivative curvature couplings $F_n$ in both
theories.
Since the $F_n$'s depend only
on the vector multiplets they are independent  on the type II
dilaton and one finds that they  arise solely at the $n$-loop order.  In the
large radius limit
one obtains  \cite{BCOV:ell,BCOV:big}
\begin{equation}
\eqalign{
F_1 &= - {4\pi i \over 12}\, \sum_\a (D_\a\cdot c_2)\, t_\a
        +\  {\rm worldsheet\ instantons}\ ,  \cr
F_{n>1} &= \  {\rm const.}\ +\  {\rm  worldsheet\ instantons}\ , }
     \label{eq:Fgtt}
\end{equation}
where $c_2$ denotes the second Chern class of $X$.
Furthermore, $g^{-2}_1$ develops a holomorphic anomaly
given by \cite{BCOV:ell}
\begin{equation}
\del_\a \bar{\del}_{\bar \b}\ g^{-2}_1
= \ff16 (17 + 5 \nv + \nh)\, \del_\a \bar{\del}_{\bar \b} K
  -  R_{\a\bar \b}\ ,
     \label{eq:anomalyII}
\end{equation}
where $R_{\a\bar \b}$ is the Ricci-tensor on the moduli space.
Let us first
note that for $n>1$ the large $S$ limits of (\ref{eq:Fnhettree}) and
(\ref{eq:Fgtt}) already agree with no further conditions
imposed. However, identifying the $F_1$'s of the dual pairs results in
a constraint on $c_2(X)$.  Note that the holomorphic anomaly of
(\ref{eq:anomalyII}) simplifies if, in the large $J$ limit, the type II
prepotential and K\"ahler potential are identical with the heterotic
$F_0$ and $K$ of (\ref{eq:Fhettree}) and (\ref{eq:Kahler}).
A straightforward computation
shows that (\ref{eq:anomalyII}) reduces in this limit  to
\begin{equation}
\del_\a \bar{\del}_{\bar \b}\ g^{-2}_1
= {b\over 12}\,  \del_\a \bar{\del}_{\bar \b} K \ ,
     \label{eq:anomalyIIb}
\end{equation}
where $b$ is defined in (\ref{eq:anomaly}).
We see that the anomalies of a dual pair coincide up to the overall
normalization.
In turn this can be used to align  the overall normalization between
 the $F_1$'s in each vacuum. Identifying the  $F_1$'s  using
(\ref{eq:ax-II}), (\ref{eq:Fnhettree}), (\ref{eq:anomaly}),
(\ref{eq:anomalyIIb})
and (\ref{eq:Fgtt}) results in
\begin{equation}
  D_s\cdot c_2(X)= 24.
     \label{eq:ctwo}
\end{equation}

We may now use the results of Oguiso \cite{Og:K3f} which state that
\begin{theorem}
Let $X$ be a minimal \CY\ threefold. Let $D$ be a nef divisor on
$X$. If the numerical $D$-dimension of $D$ equals one and $D\cdot c_2(X)>0$
then there is a fibration $\Phi:X\to W$, where $W$ is $\P^1$ and the
generic fibre is a K3 surface.
\end{theorem}
We have obtained the desired result --- $X$ has to be  a
K3-fibration in order to be the dual partner of a heterotic vacuum.
Note that in general the fibre is allowed to degenerate
to something other than a K3 surface over a finite number of points on
$W$. This fibration is shown in figure \ref{fig:fib}.

\iffigs
\begin{figure}
  \centerline{\epsfxsize=10cm\epsfbox{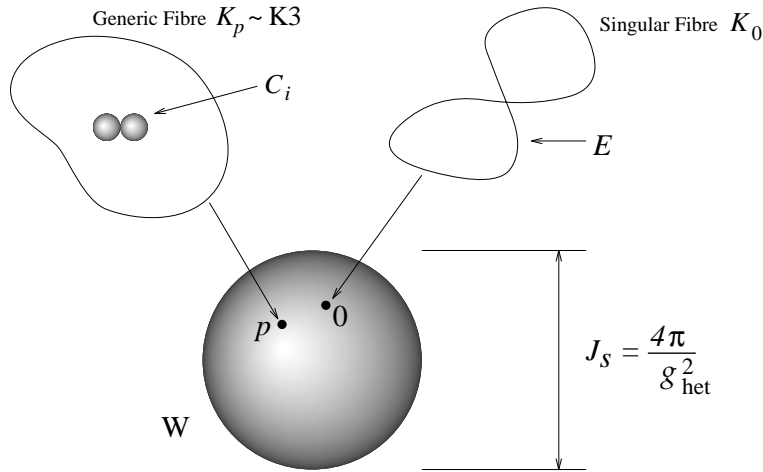}}
  \caption{$X$ as a K3-fibration.}
  \label{fig:fib}
\end{figure}
\fi

We should emphasize two points about our derivation of the
result. Firstly, we assumed that the \CY\ manifold in the type IIA
theory was in the \CY\ phase when the heterotic string was
weakly-coupled. It is certainly conceivable that there may be dual
pairs where the phases are not aligned like this. In such an example,
there is then no reason why a \CY\ manifold on the type IIA side, which
appears for some strongly-coupled regime of the heterotic moduli
space, should be a K3-fibration. In other words, the weakly-coupled
heterotic string might be dual to a type IIA string theory on a manifold
$X$, which is not a K3-fibration, in the following sense. The instanton sum in
(\ref{eq:Ftt}) is necessarily divergent but could be reinterpreted
in terms of some other non-\CY\ phase (such as a perturbed
Landau-Ginzburg theory or orbifold \cite{AGM:II}) to give a convergent
sum of corrections.

Secondly, we are stating only that the type IIA string is compactified on a
K3-fibration. The type IIB string is compactified on $Y$, the mirror
of $X$, which may or may not be a K3-fibration. The question as to
what extent mirror symmetry acts within the class of K3-fibrations has
not been analyzed at this point in time so we cannot make any strong
statements about when $Y$ is a K3-fibration. One can observe though
that in the
cases discussed in \cite{KV:N=2}, where the mirror
map is in the Greene-Plesser orbifold form \cite{GP:orb}, then the
mirror orbifolding operation preserves the fibration structure and so both
$X$ and $Y$ are K3-fibrations.

Note that (\ref{eq:ctwo}) not only tells us that $D_s\cdot c_2(X)>0$ but
that the value is the Euler characteristic of a K3 surface. This tells
us that the element in $H_4(X)$ corresponding to $D_s$ is given
precisely by a generic K3 fibre \cite{Og:K3f}. Clearly the 2-cycle
dual to this is
given by $W$, the base space. It then follows that $J_s$ is the
cohomology element that gives us the {\em size\/} of this
2-cycle. That is, the dilaton in the heterotic string is given by the
size of the base $\P^1$ in the fibration of $X$.

Thus far we have made no appeal to string-string duality in
six-dimensions but we can use this duality to reproduce the result we
have just found concerning the dilaton. Let us set the axion in the
heterotic string to zero for simplicity. In \cite{W:dyn} it was shown
that
\begin{equation}
  S_6 = {1\over S_6^\prime}\ , \qquad
           S_6 h_6 = h_6^\prime,
\end{equation}
where $S_6$ is the dilaton in six dimensions for the heterotic string
and $h_6$ is its metric with the corresponding values primed for the
type IIA string. Let us apply this to our picture in four dimensions
by ``compactifying'' the six-dimensional space-time over $\P^1$. Given
the Weyl-rescaling between the type IIA and heterotic space-times, the
areas of these rational curves will differ by a factor of $S_6$.
The action for the heterotic string can then be written
\begin{equation}
  \eqalign{L_{\rm het} &= \int d^6x\, \sqrt{h_6}S_6(R+\ldots)
  = \Area(\P^1)\int d^4x \, \sqrt{h_4}S_6(R+\ldots)\cr
  &= \Area(\P^1)^\prime \int d^4x \, \sqrt{h_4}(R+\ldots).\cr}
\end{equation}
This last form of the action tells us that the area of the base in the
type IIA picture gives the four-dimensional dilaton of the heterotic
string as we saw above.

For the dilaton modulus in the type IIB picture we can use the
monomial-divisor mirror of \cite{AGM:mdmm} (or, more specifically, the
results of \cite{CDFKM:I,HKTY:}) to find the corresponding
deformation of complex structure of $Y$. Note also
that we know the result exactly --- since the dilaton is given by the
K\"ahler form, the map between the dilaton and the complex modulus of
$Y$ is corrected by solutions to the Picard-Fuchs equation.

For example, in \cite{KV:N=2} $Y$ was taken to be an orbifold of the
hypersurface
\begin{equation}
  x_1^{12}+x_2^{12}+x_3^6+x_4^6+x_5^2-12\psi x_1x_2x_3x_4x_5 -
        2\phi x_1^6x_2^6=0,
\end{equation}
in $\P^4_{\{1,1,2,2,6\}}$. It was then conjectured that there was a
heterotic string dual to a type IIB string compactified on $Y$. From
what we have said above and using the results of \cite{CDFKM:I} it
follows that the heterotic dilaton is given by
\begin{equation}
  S = -\frac1{8\pi^2}\left( \log(z_1)+2z_1+240z_2+3z_1^2-480z_1z_2
    +220680z_2^2+\ldots\right),         \label{eq:eg1}
\end{equation}
where
\begin{equation}
  z_1 = \frac1{4\phi^2}\ , \qquad
  z_2 = -\frac{2\phi}{(12\psi)^6}.
\end{equation}
This agrees with the conjectured relationship in \cite{KV:N=2}. One
can show that the example with $h^{2,1}(Y)=3$ of \cite{KV:N=2} is also
consistent with our statements. It is perhaps worth emphasizing that
we have deduced (\ref{eq:eg1}) given only the
existence of the weakly-coupled heterotic string --- no further
knowledge of the heterotic string (such as enhanced gauge symmetries)
was required.

Let us now turn  to the other $\nv-1$ vector moduli.
Now that we know that $X$ is a K3-fibration, we
can say exactly where the other contributions to $h^{1,1}(X)$ come
from. The general statement is that they can be divided into two
classes --- those from the generic fibre and those from the singular
fibre. Let us concentrate first on the contribution from the generic
fibre.

Let $C_i$ be an algebraic curve in a generic K3 fibre. As we move
about the base space we can map the curve into equivalent curves in
the other generic fibres. Adding all these curves together gives us a
divisor, $D_i$ in $X$. That is, we have written $D_i$ as a fibration
over $W$ where $C_i$ is the generic fibre. (Note that $D_i$ is not
dual to $C_i$.) Thus, a curve in the generic fibre contributes to
$h^{1,1}$. Because a generic K3-fibration has monodromy, we may have
two curves in a generic fibre that can be mapped into each other by
cycling round a path on $W$. Clearly such a pair only contribute one
to $h^{1,1}$ as they both were required in building $D_i$.

Let us denote a generic fibre $K_p$, where $p\in W$. The set of (duals
of) algebraic curves in $K_p$ generate a sublattice of $H^2(K_p,\Z)$
called the ``Picard lattice'' of $K_p$. Clearly both $H^{1,1}(K_p)$
and the lattice $H^2(K_p,\Z)$ can be embedded in $H^2(K_p,\C)$. The
Picard lattice can be considered the intersection $H^{1,1}(K_p)\cap
H^2(K_p,\Z)$.

Consider the space $H^2(K_p,\R)$. The natural wedge product between
elements in this space endows it with a metric.
One can show that this
metric has signature (3,19) (see, for example, \cite{BPV:}).
Let $\omega$ be a $(2,0)$-form on $K_p$. One can show that the real
and imaginary parts of $\omega$ span a space-like 2-plane in
$H^2(K_p,\R)\cong\R^{3,19}$. As any element of the Picard lattice is a
(1,1)-form and so lies orthogonal to this 2-plane, the Picard lattice
must be embeddable in $\R^{1,19}$. The ``Picard number'' of a K3
surface is the rank of the Picard lattice. Clearly then, the Picard number
can be no bigger than 20.

The K\"ahler form on a K3 surface is a vector within the real space generated
by the Picard lattice. The volume of the K3 surface is given by the
length of this vector which must therefore be positive. Thus we have
shown that the Picard lattice must have signature $(+,-,-,\ldots,-)$.
{}From what we have said above, the vector multiplets coming from the
generic fibre are the generators of the monodromy-invariant subspace
of the Picard lattice. This subspace must contain the K\"ahler
form of the fibre and so also has signature $(+,-,-,\ldots,-)$. The
rank is clearly less than or equal to the Picard number.

Consider the intersection numbers $D_s\cdot D_i\cdot D_j$ where the latter two
divisors come from curves in the generic fibres. This is equal to the
intersection $C_i\cdot C_j$ within the fibre and is thus given by the
intersection
form of the monodromy-invariant Picard lattice. This means that we
have exactly reproduced the heterotic string result
(\ref{eq:Fhettree}) in terms of the type IIA string.

We also know that there is at least one (the volume of the fibre)
but no more than 20  moduli from this
source. Adding these to the dilaton and the $U(1)$ from the
gravity multiplet shows that the rank of the gauge group
for K3-fibrations with only  contributions to $h^{1,1}$ from the
base space and generic fibers satisfies
\begin{equation}
3 \leq  \rank(G_{II}) \leq 22\ ,    \label{eq:IIrank}
\end{equation}
which is consistent with the heterotic bound (\ref{eq:hetrank}).
The upper limits  are suspiciously close to each other
and  we can probably bring these results into
agreement once we take into account the {\em quantum\/} geometry of
the K3 fibre. In the above analysis we used purely classical geometry
for the analysis of the Picard lattice. Given the appearance of
quantum effects in K3 surfaces \cite{AM:K3p} it is not at all
unreasonable to expect that some quantum Picard lattice exists for
a Planck-sized K3 surface with Picard number 22.
This would bring the upper limit of $\GII$
 into complete agreement with the heterotic limit.
The lower limit of the heterotic vacuum
can never be reached for a Calabi--Yau manifold since
there necessarily has to be a volume form $J$ which obeys
$J\cdot J\cdot J\neq 0$ and thus cannot be the image of the heterotic
dilaton.
However it is conceivable that a model which does not have a \CY\
phase may be used to achieve this lower limit.


\section{Limitations of duality} \label{s:fail}

So far we have discussed the successes of the type IIA/heterotic
duality picture. Now let us discuss some of the short-comings and the
reasons why they must appear.

The first question which comes to mind concerns the contributions to
$h^{1,1}(X)$ from the degenerate fibres. Let us call such a divisor
$E$. Clearly such a divisor will not intersect the generic fibre $D_s$
and so
\begin{equation}
  D_s\cdot E\cdot E = 0,              \label{eq:decoup}
\end{equation}
in contradiction with (\ref{eq:Fhettree}) for the heterotic
string. This implies that $K3$-fibrations where $h^{1,1}(X)$
contains  moduli corresponding
to degenerate fibers cannot be the dual of a standard weakly coupled
heterotic string vacuum as we described it in section~2.
However, such fibrations  still  have a candidate
for the image of the heterotic dilaton  but
there seem to be no heterotic states which $E$ can be
the image of. If there are such states  (\ref{eq:decoup})
implies that they decouple at the tree level
and only arise at one-loop and/or nonperturbatively
(equivalently this says  that they cannot be states in the conformal
field theory).
Since in general $E\cdot E\cdot E \neq 0$ the corresponding  heterotic
states  would have  to arise already
at the one-loop level.
Alternatively  they   could  be states that become massless only
nonperturbatively  but
with ``unusual'' couplings to the dilaton such as the
Ramond-Ramond states  in type II strings.
Being members of vector multiplets they would  enhance
the rank of  $\Ghet$ nonperturbatively but
so far no such states have been identified directly
in the heterotic string.

The number of states coming from singular fibres can be very large
easily taking the rank of the gauge group beyond the expected limit of
24.  But as we have just argued  these vector multiplets
have no conformal field theory interpretations on the heterotic
side and thus  no contradiction arises.  Conversely, demanding the absence
of states corresponding to degenerate fibers implies the bound
(\ref{eq:IIrank})
perfectly consistent with the heterotic string.

What if $X$ is not a K3-fibration? What happens to the heterotic dual?
This question is remarkably similar to the question of if a manifold
has a mirror partner. A simple example of a manifold without a mirror
partner is that of one with no deformations. This implies $h^{2,1}=0$
and so the mirror would have $h^{1,1}=0$ which is impossible for a
K\"ahler manifold. It turns out however that even in many cases where
$h^{2,1}>0$, the manifold may have no mirror. This was analyzed in
\cite{AG:gmi} where it was shown that some $N=2$ superconformal field
theories, appearing as the mirror of a \CY\ theory, have a non-trivial
moduli space of (1,1)-forms but nowhere in
this moduli space is there a \CY\ phase.\footnote{It is also possible
that discrete torsion can lead to similar effects \cite{Berg:dt}.}

We claim that the same thing happens for type II/heterotic
duality. The moduli space of vector multiplets in a type IIA string
may be non-trivial but
it may be that there is simply no place in this moduli space where
there is a weakly-coupled heterotic string. This should happen if $X$
is not a K3-fibration. Of course, this works both ways. Given the
moduli space of vector multiplets of a heterotic string there may be
no place in this space where a weakly-coupled type IIA string
compactified on a \CY\ manifold can be described.

It is worth emphasizing that missing dual partners can enter into the
picture of connecting up the moduli space of $N=2$ theories via phase
transitions as in \cite{GMS:con,me:flower} and flops as in
\cite{AGM:I}. It may be that there is an
extremal transition (such as a conifold transition) or flop\footnote{
We thank  P.~Berglund for discussions on this point.}
on the type IIA
side from a manifold which is a K3-fibration to a manifold which is
not. In this case, the extremal transition or flop
involves shrinking down the base $\P^1$ of the fibration. This means
that  on the heterotic side the phase transition occurs at  strong coupling
and the ``new'' phase no longer has a (heterotic) weak coupling limit.


\section*{Acknowledgements}

We thank M.~Gross for many important contributions. It is also a
pleasure to thank P.~Argyres, P.~Berglund, S.~Chaudhuri, B.~de Wit, B.~Greene,
J.~Harvey, V.~Kaplunovsky, D.~Morrison, J.~Polchinski, R.~Schimmrigk
and A.~Strominger  for useful conversations as well as all the other
``Plancksters'' at the I.T.P., Santa Barbara program where part of
this work was done under the NSF grant PHY94-07194. We are especially
grateful to P.~Langacker, P.~Nath and J.~Polchinski for organizing a
stimulating workshop. The work of P.S.A.~is supported by a grant from
the National Science Foundation.
The work of J.L.~is supported by a Heisenberg fellowship
of the DFG and the German-Israeli foundation.

\end{document}